\documentclass[openacc]{rstransa}%%%%where rstrans is the template name

\usepackage{graphicx, epsfig}
\usepackage{color}
\usepackage{mathrsfs}
\usepackage{bm}
\usepackage{amsmath,amssymb}% for \eqref
\usepackage[caption=false]{subfig}

\newcommand{\be}{\begin{equation}}
\newcommand{\ee}{\end{equation}}
\newcommand{\ba}{\begin{eqnarray}}
\newcommand{\ea}{\end{eqnarray}}
\newcommand{\bpm}{\begin{pmatrix}}
\newcommand{\epm}{\end{pmatrix}}

\newcommand{\mmbar}{$M{\bar M}$~}

\begin{document}

%%%% Article title to be placed here
\title{Monopole-antimonopole: interaction, scattering and creation}

\author{%%%% Author details
Ayush Saurabh, Tanmay Vachaspati}

%%%%%%%%% Insert author address here
\address{$^{1}$Physics Department, Arizona State University, Tempe, AZ 85287, USA.
}

%%%% Subject entries to be placed here %%%%
\subject{field theory, particle physics}

%%%% Keyword entries to be placed here %%%%

%%%% Insert corresponding author and its email address}
\corres{Insert corresponding author name\\
\email{tvachasp@asu.edu}}

%%%% Abstract text to be placed here %%%%%%%%%%%%
\begin{abstract}
The interaction of a magnetic monopole-antimonopole pair depends on
their separation as well as on a second ``twist'' degree of freedom. This
novel interaction leads to a non-trivial bound state solution known as
a sphaleron and to scattering in which the monopole-antimonopole
bounce off each other and do not annihilate. The twist degree of
freedom also plays a role in numerical experiments in which gauge
waves collide and create monopole-antimonopole pairs. Similar
gauge wavepacket scatterings in the Abelian-Higgs model lead to
the production of string loops that may be relevant to superconductors.
Ongoing numerical experiments to study the production of electroweak
sphalerons that result in changes in the Chern-Simons number, and
hence baryon number, are also described but have not yet met with success.
\end{abstract}
%%%%%%%%%%%%%%%%%%%%%%%%%%%

%%%%%%%%%% Insert the texts which can accomdate on firstpage in the tag "fmtext" %%%%%

\maketitle

\section{Introduction}
\label{introduction}

Magnetic monopoles have been known for over 40 years now as regular solutions 
in non-Abelian gauge theories~\cite{thooft,polyakov1974spectrum}. 
They provide a fertile playground for theoretical ideas~\cite{rebbi1985book} 
and are also relevant to physical considerations as they are necessarily present in all 
grand unified models~\cite{Preskill:1986kp}. In the standard 
model of the electroweak interactions, only confined magnetic monopoles exist.  
Even so, they
can lead to insight into processes such as baryon number violation~\cite{vachaspati1994electroweak}.

In the present work we are interested in properties of monopole-antimonopole (\mmbar) 
pairs. How do monopoles interact with antimonopoles when they are at rest? How do they
scatter? Can they be produced in scattering experiments? Can they play a role in particle
physics?
There is a large body of work
on the first two questions and there are also several excellent texts~\cite{rebbi1985book,manton2007book} 
where the reader can access results.  Our recent work~\cite{tanmaycreation, tanmayscattering, 
saurabh2017interaction} provides numerical evidence for some of these works and extends them 
in some cases.  The question of \mmbar creation has also received attention but it is 
difficult to answer
especially as it seems to require a description of a non-perturbative final state (\mmbar)
in terms of perturbative initial states (particles). The non-perturbative state is best described
in classical terms (solutions of certain differential equations) while the perturbative
state is best described in terms of quanta in a quantum field theory. Thus the 
process also requires a description that enables transition from quantum to classical
variables. We shall largely bypass these deep questions and study the creation of
\mmbar when the initial state has large occupation number and can be described
in classical terms. After all, the initial state is up to us to prepare and we are free
to set it up as we wish.

There are two applications of the work on \mmbar creation that are more immediate.
The first is that just as we can consider the creation of \mmbar, we can also
consider the creation of string loops. Indeed we find initial conditions in the Abelian-Higgs
model that lead to string creation. These strings are produced in loops that live for
a short time and then collapse. In some regime of parameters, the Abelian-Higgs
model also provides a description of superconductors, leading to the possibility
of experimentally producing strings in superconductors. (This is similar to the
production of string loops in He-3 by the bombardment of neutrons in an 
experiment that has already been done~\cite{ruutu1996vortex}.) The second application 
of the work on \mmbar creation is in the context of the electroweak model. Here
monopoles are confined and a monopole is always connected to an antimonopole
by a Z-string. If we are able to create a confined electroweak \mmbar, it will re-annihilate
just as the loops of string in the Abelian-Higgs model re-collapse. Further, if the
electroweak \mmbar annihilates after some specific dynamics, the electroweak 
Chern-Simons number can change and lead to baryon number violation.  Thus a 
better understanding of the creation of monopoles may enable processes in which 
baryon (and/or lepton) number changes. However, ongoing numerical work on the 
production of electroweak \mmbar has not yet resulted in a change of Chern-Simons 
number.

\section{$M{\bar M}$ in SO(3) model}
\label{mmbarso3}

We consider an SO(3) gauge theory with a scalar in the adjoint representation with Lagrangian
\be
\mathcal{L}=\frac{1}{2}(D^{\mu}\phi)^{a}(D_{\mu}\phi)^{a}-\frac{1}{4}W^{a\mu\nu}W_{\mu\nu}^{a}
-\frac{\lambda}{4}(\phi^{a}\phi^{a}-\eta^{2})^{2}
\ee
where $a=1,2,3$, the covariant derivative is defined as,
\be
(D_{\mu}\phi)^{a}=\partial_{\mu}\phi^{a}+g\epsilon^{abc}W_{\mu}^{b}\phi^{c}
\ee
and the gauge field strength is given as 
\be
W_{\mu\nu}^{a}=\partial_{\mu}W_{\nu}^{a}-\partial_{\nu}W_{\mu}^{a}+g\epsilon^{abc}W_{\mu}^{b}W_{\nu}^{c}
\ee
The energy of a static field configuration is given by,
\be
E=\int d^{3}x \left [\frac{1}{2}(D_{i}\phi)^{a}(D_{i}\phi)^{a}+\frac{1}{4}W_{ij}^{a} W_{ij}^{a}
+\frac{\lambda}{4}(\phi^{a}\phi^{a}-\eta^{2})^{2} \right ]
\label{energydensity}
\ee

It is known~\cite{thooft} that the model has a monopole solution
\be
\phi^{a}=h(r)\hat{r^{a}}, \ \ W_{i}^{a}=\frac{(1-k(r))}{r}\epsilon^{aij}\hat{r}^{j}
\ee
where $h$ and $k$ are profile functions that can be found by solving the
equations of motion with the boundary conditions
\be
h(0 )=0,\ k(0)=1, h(\infty )=1,\ k(\infty)=0 .
\ee

\begin{figure}
\begin{center}
\includegraphics[height=0.35\textwidth,angle=0]{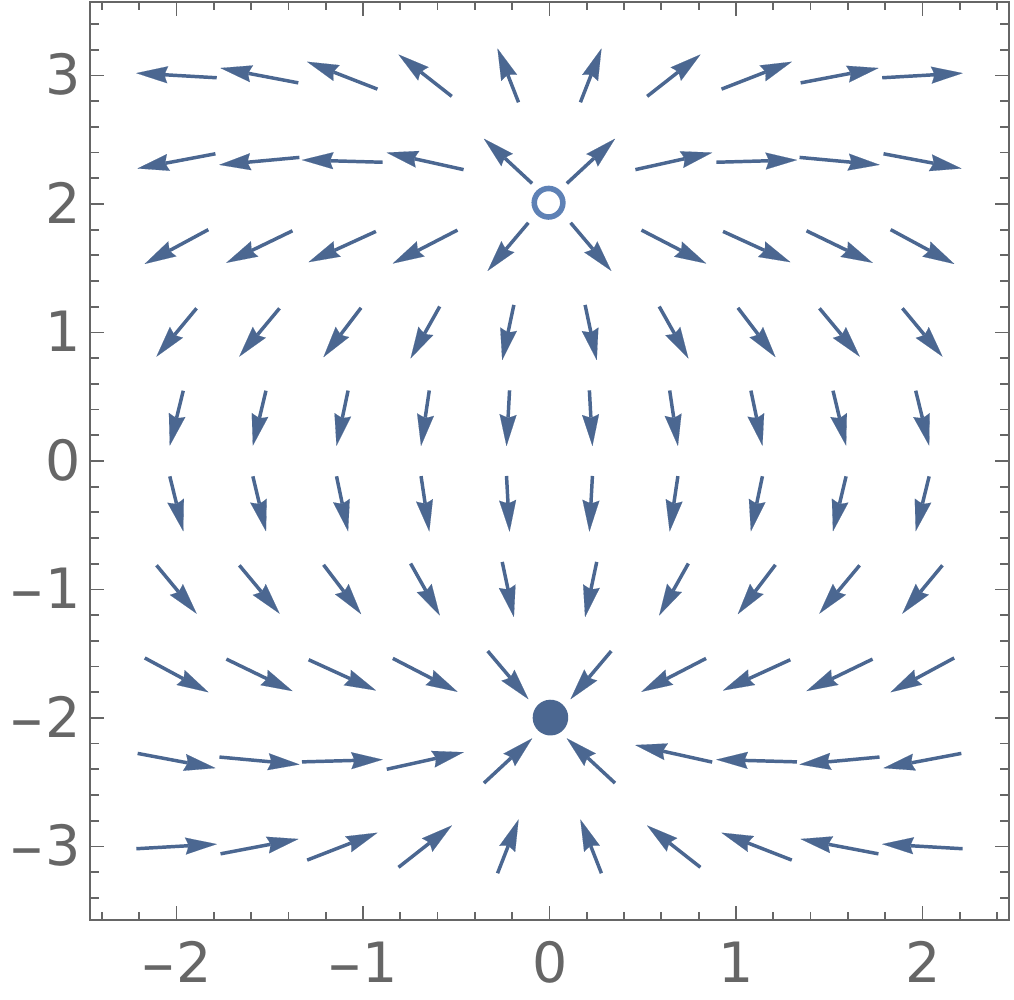}
\includegraphics[height=0.35\textwidth,angle=0]{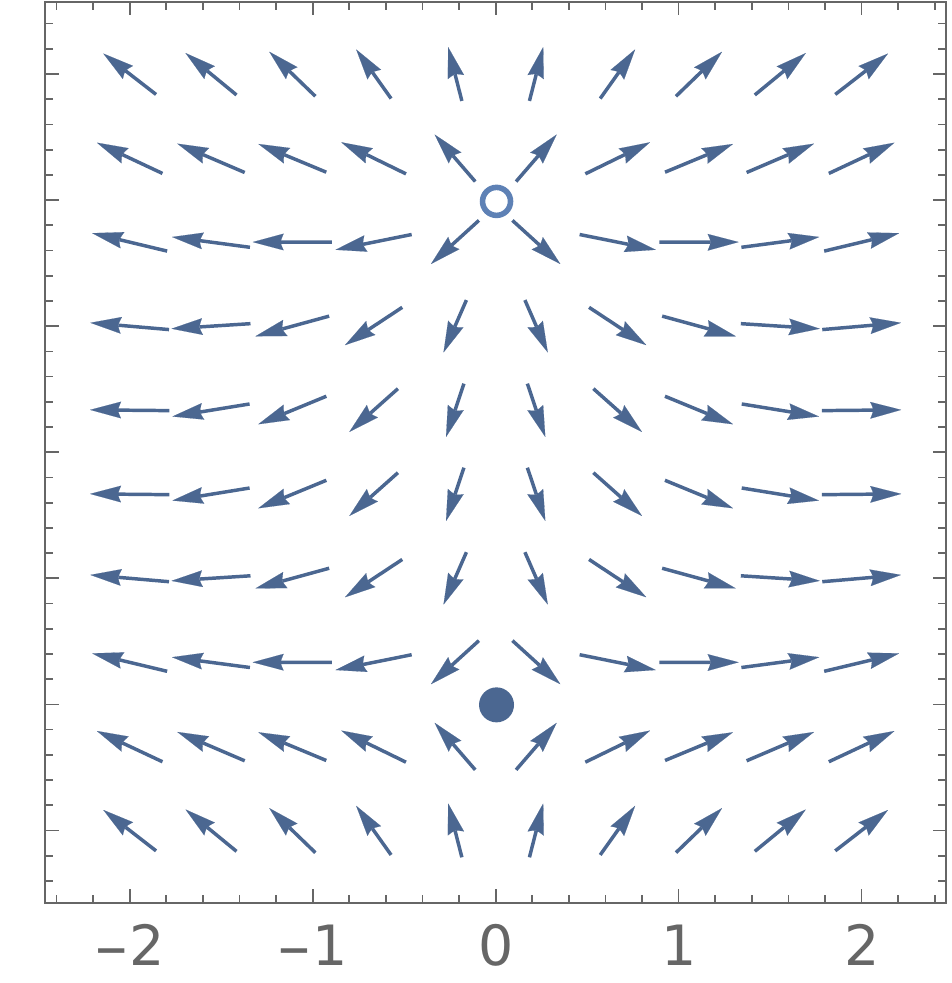}
\end{center}
\caption{Higgs vectors in the $xz-$plane for twist = $\pi$ (left) and twist = $0$ (right). 
The Higgs zeros are located at $(0,2)$ and $(0,-2)$, shown as filled and unfilled circles.
}
\label{gaugePi}
\end{figure}

The \mmbar configuration can now be written by gluing together a monopole and an antimonopole.
There is some freedom in this procedure but all we need is that the monopole and antimonopole
be located with some fixed separation and that they should have a fixed relative twist. Then the 
fields can be relaxed to the lowest energy configuration subject to these constraints, and/or used 
as initial conditions for time evolution. We choose the \mmbar scalar field orientations to be
given by
\ba
\hat{\phi}^{1}&=&(\sin\theta \cos\bar{\theta} \cos \gamma-\sin\bar{\theta}\cos\theta) \cos(\varphi-\gamma/2) 
 -\sin\theta \sin\gamma \sin(\varphi-\gamma/2)
 \label{phi1} \\
\hat{\phi}^{2}&=&(\sin\theta \cos\bar{\theta}\cos\gamma-\sin\bar{\theta}\cos\theta)\cos(\varphi-\gamma/2)
-\sin\theta \sin\gamma \cos(\varphi-\gamma/2)
\label{phi2} \\
\hat{\phi}^{3}&=&\cos\theta \cos\bar{\theta}+\sin\theta \sin\bar{\theta}\cos\gamma
\label{phi3}
\ea
Note that the configuration has two free parameters. The separation of the monopole and antimonopole, $d$,
is hidden in the spherical angles $(\theta,\phi)$ and $({\bar \theta},{\bar \phi})$ that are with respect to
coordinate systems with origins at the monopole and antimonopole respectively. The second is the
twist paramter $\gamma$. The scalar field configurations for $\gamma=0,\pi$ are illustrated in 
Fig.~\ref{gaugePi}. The expression for the scalar field with the profile functions included is
\be
\phi^a = h(r_m) h(r_{\bar{m}}) {\hat \phi}^a
\label{phistart}
\ee
where $r_m$ anad $r_{\bar{m}}$ are the distances to the monopole and antimonopole
respectively.
For the gauge fields we take the configuration,
\be
W_{\mu}^{a}=-(1-k(r_{m}))(1-k(r_{\bar{m}}))\epsilon^{abc}\hat{\phi}^{b}\partial_{\mu}\hat{\phi}^{c}
\label{initialW}
\ee

\begin{figure}
\begin{center}
  \includegraphics[height=0.30\textwidth,angle=0]{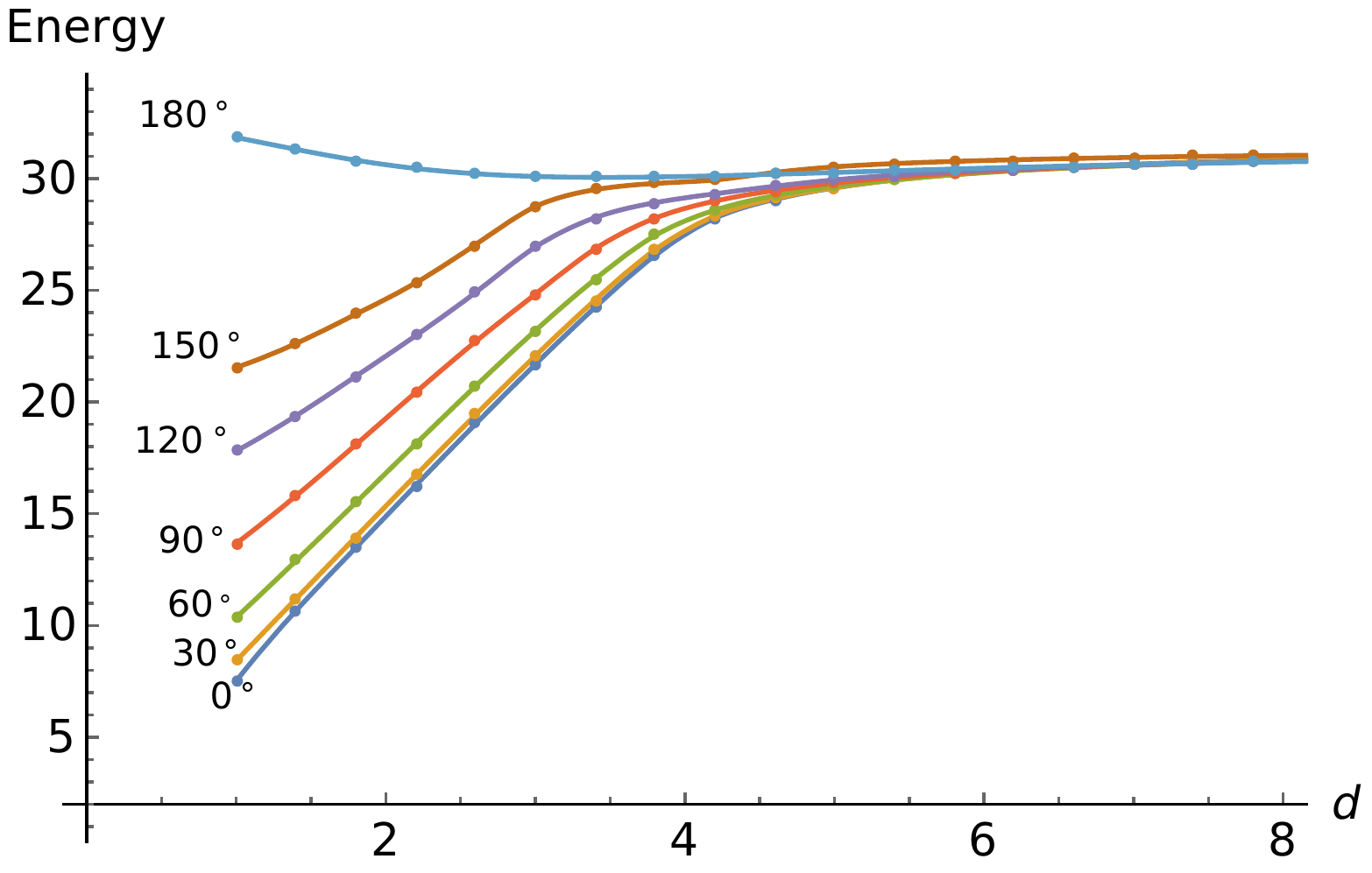}
  \includegraphics[height=0.30\textwidth,angle=0]{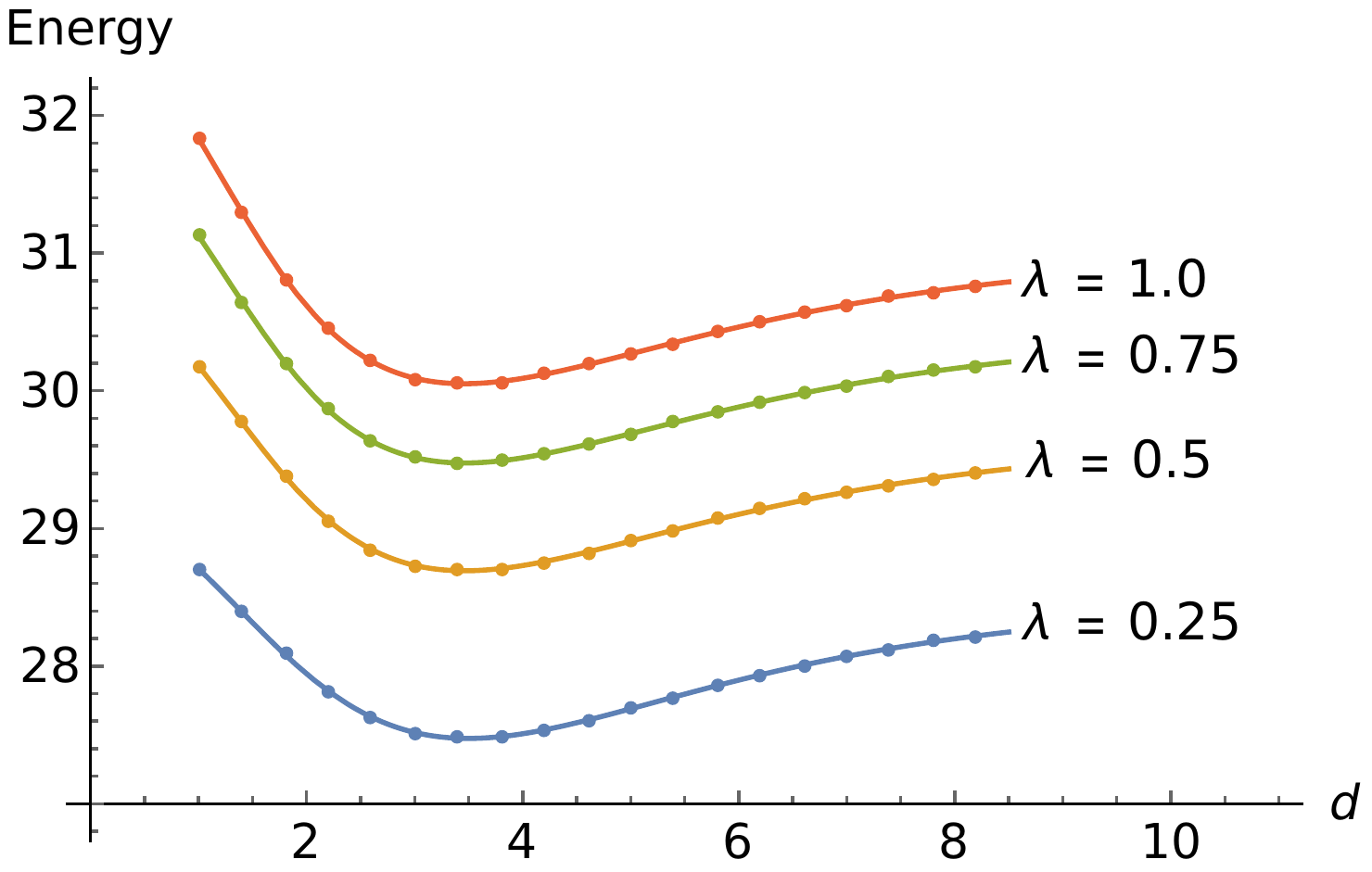}
\end{center}
  \caption{ {(\it Left)} Total energy as a function of monopole-antimonopole separation $d$ 
  for $\lambda=1$ and twist varying from $0$ to $\pi$.
 {(\it Right)} Total energy as a function of monopole-antimonopole separation $d$ for twist 
  $\gamma = \pi$ and $\lambda$ varying from $0.25$ to $1.0$. The sphaleron solution 
  is at the minimum in every curve.
 }
\label{lambda1}
\end{figure}

\subsection{$M{\bar M}$ Interaction}
\label{mmbarinteraction}

The starting configuration in Eq.~\eqref{phistart} can now be 
relaxed so as to lower the field energy but with fixed values of $d$ and $\gamma$.
A numerical scheme was implemented for this relaxation in Ref.~\cite{saurabh2017interaction} 
and the results are shown in Fig.~\ref{lambda1}. The energy depends on
$d$ and $\gamma$. The most interesting feature is that the energy curve
for $\gamma =\pi$ has a minimum. The symmetry under $\gamma \to \pi-\gamma$
then shows that the energy function has a saddle point at $\gamma=\pi$
and $d \sim 3$ (in units in which $\eta=1$ and the vector mass and the monopole
radius are 1). Thus the SO(3) 
equations have a saddle point solution
corresponding to a bound state of a monopole-antimonopole as was
first shown to exist by Taubes in the SO(3) model in 
Ref.~\cite{Taubes:1982ie,Taubes:1982ie2} and also 
discovered in the electroweak model by Manton in Ref.~\cite{mantonforce}.

\subsection{$M{\bar M}$ Scattering}
\label{mmbarscattering}

\begin{figure}[h]
  \includegraphics[height=0.3\textwidth,angle=0]{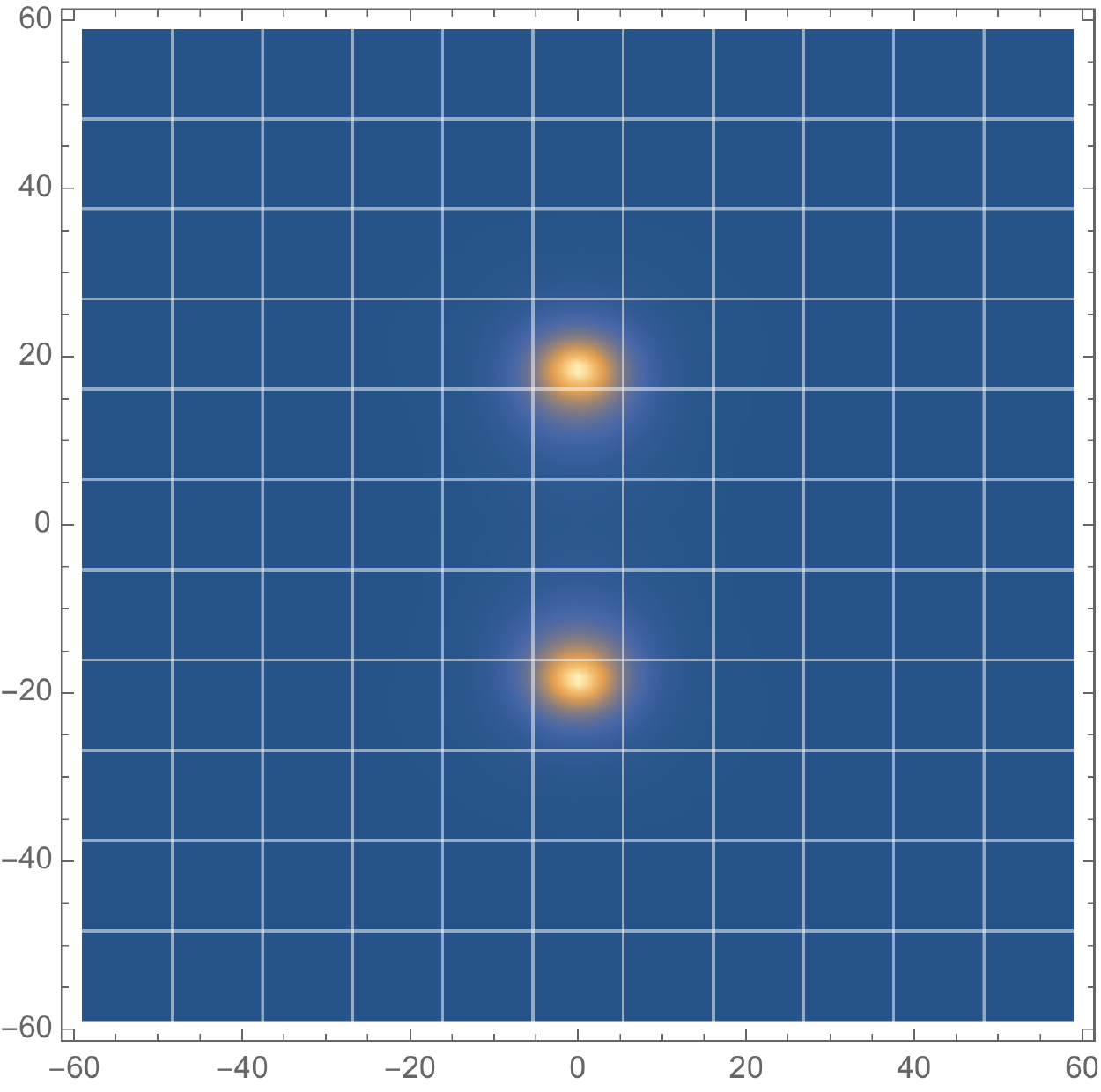}
    \includegraphics[height=0.3\textwidth,angle=0]{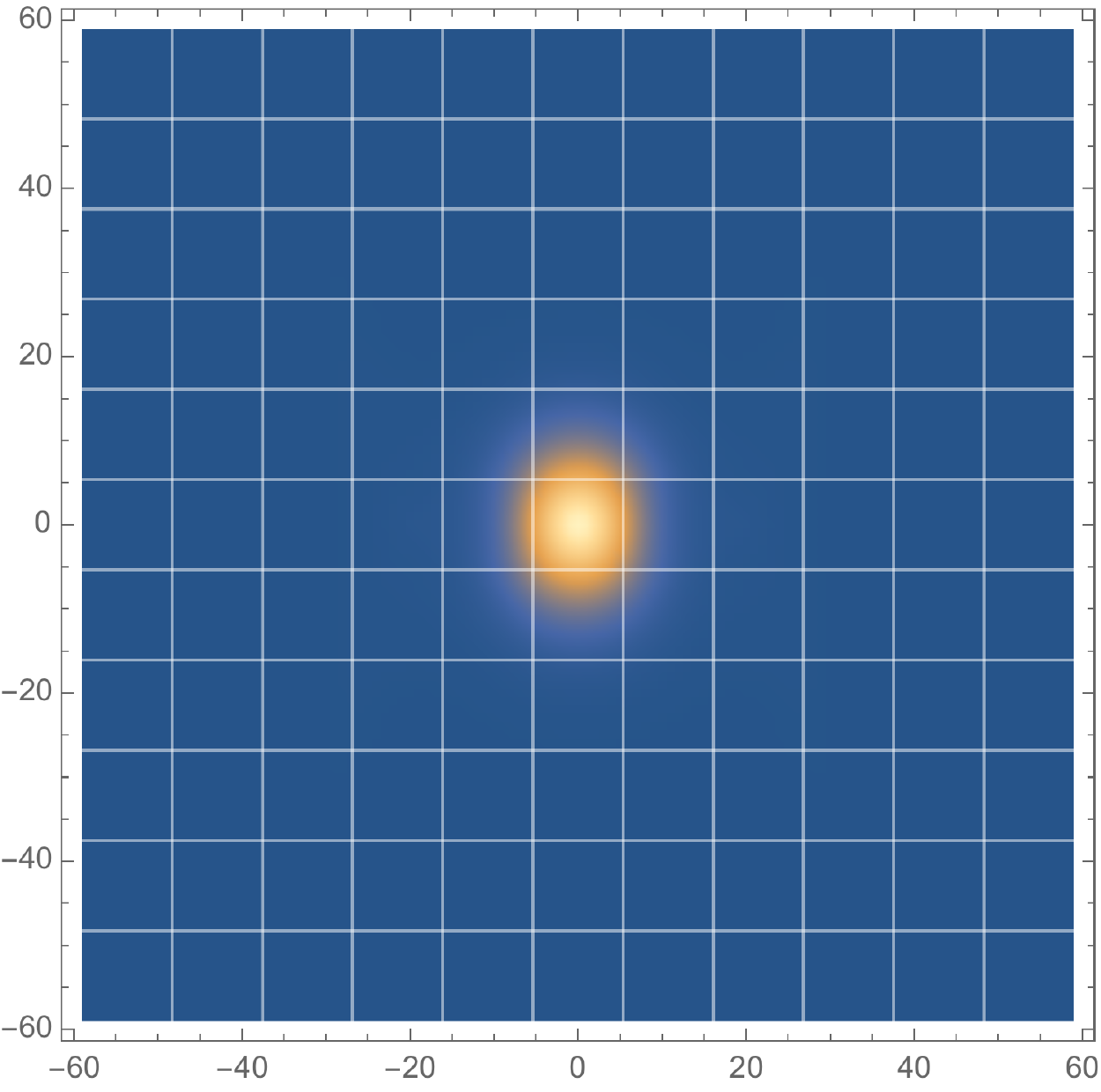}
  \includegraphics[height=0.3\textwidth,angle=0]{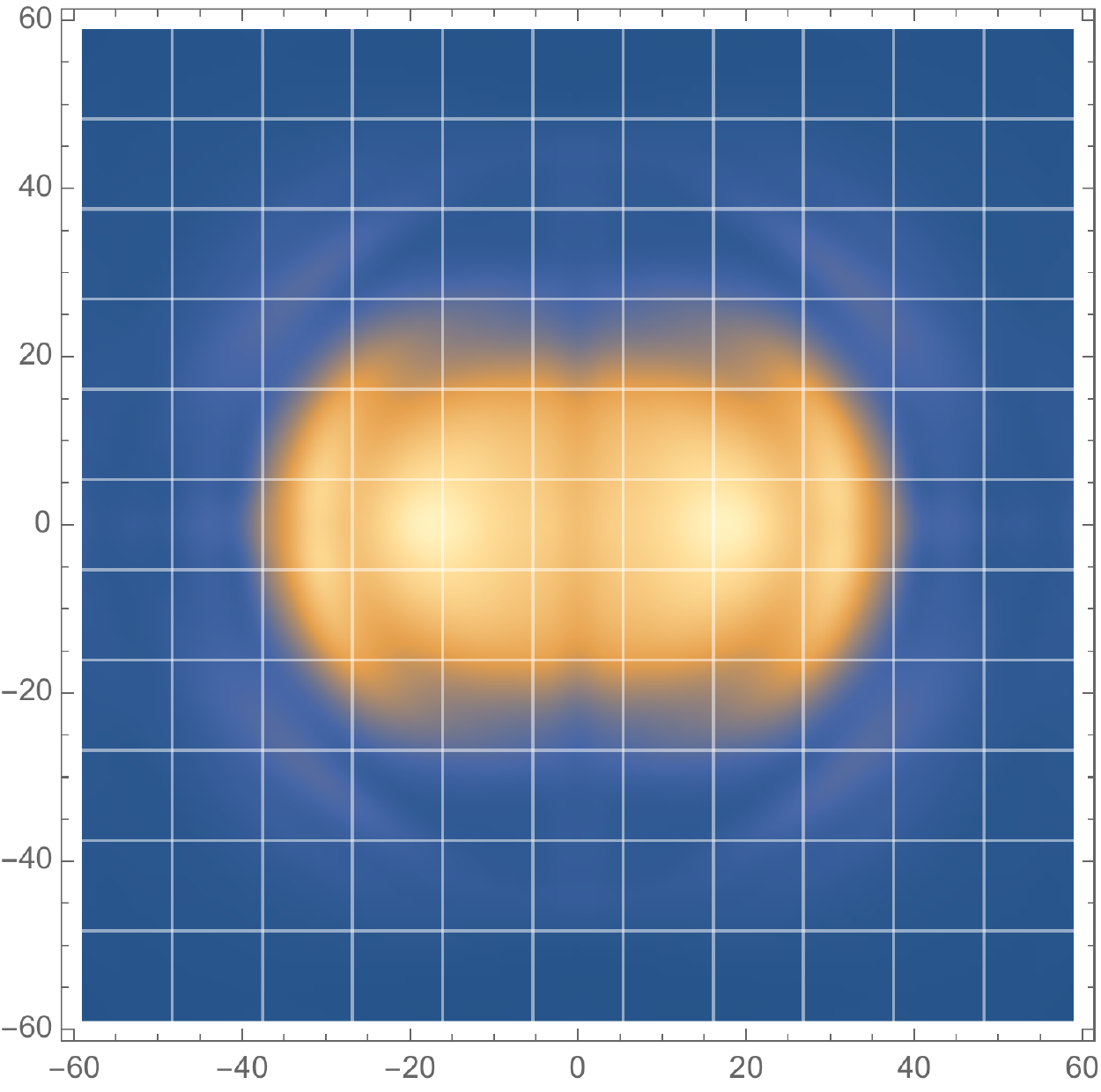}
  \caption{Snapshots of a planar slice of annihilating monopole and antimonopole for $\lambda=1$,
  $\gamma=0$, and $v_z=0.5$. The colors represent energy density.}
\label{snapshots1}
\end{figure}

\begin{figure}
  \includegraphics[height=0.3\textwidth,angle=0]{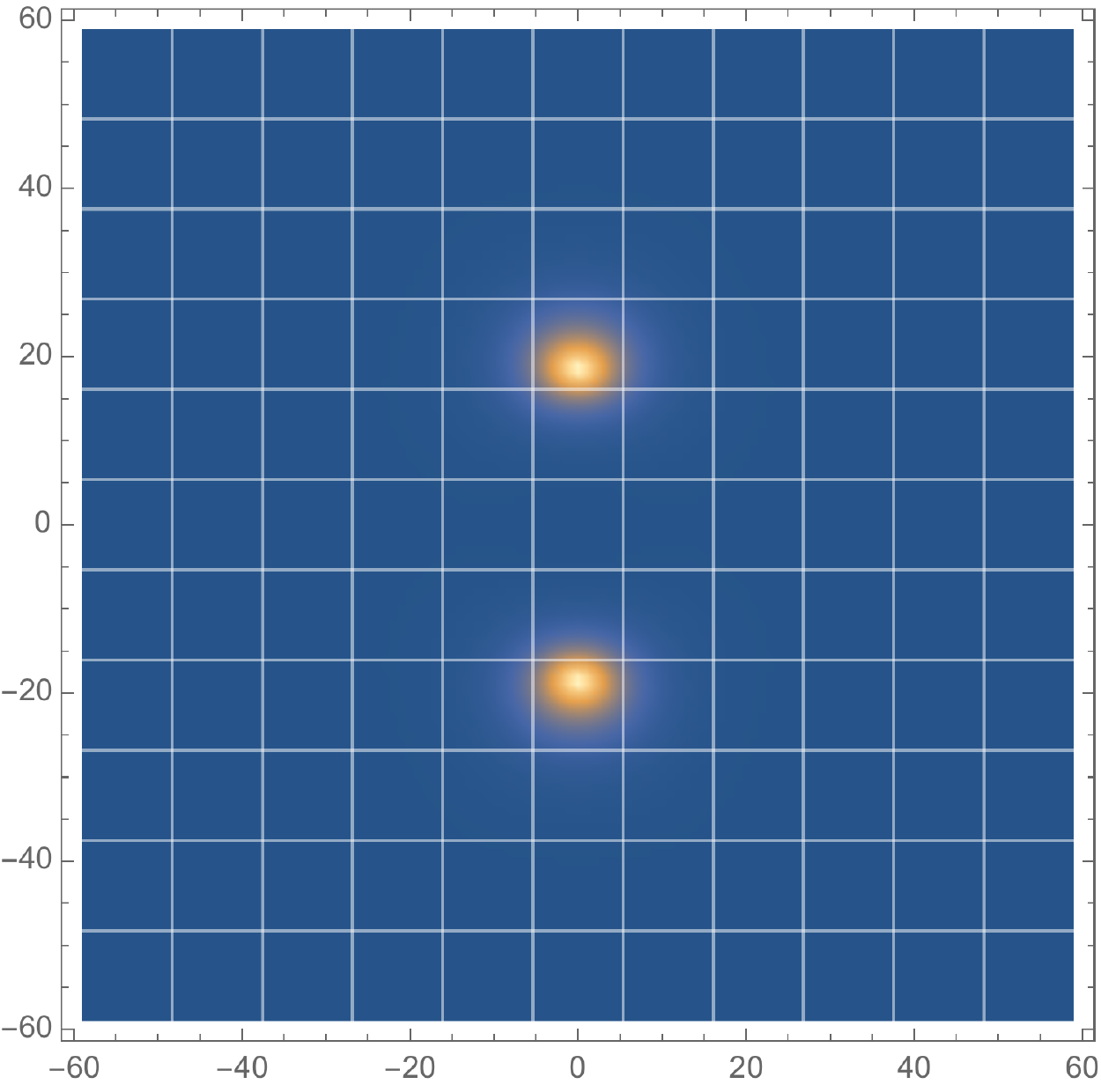}
    \includegraphics[height=0.3\textwidth,angle=0]{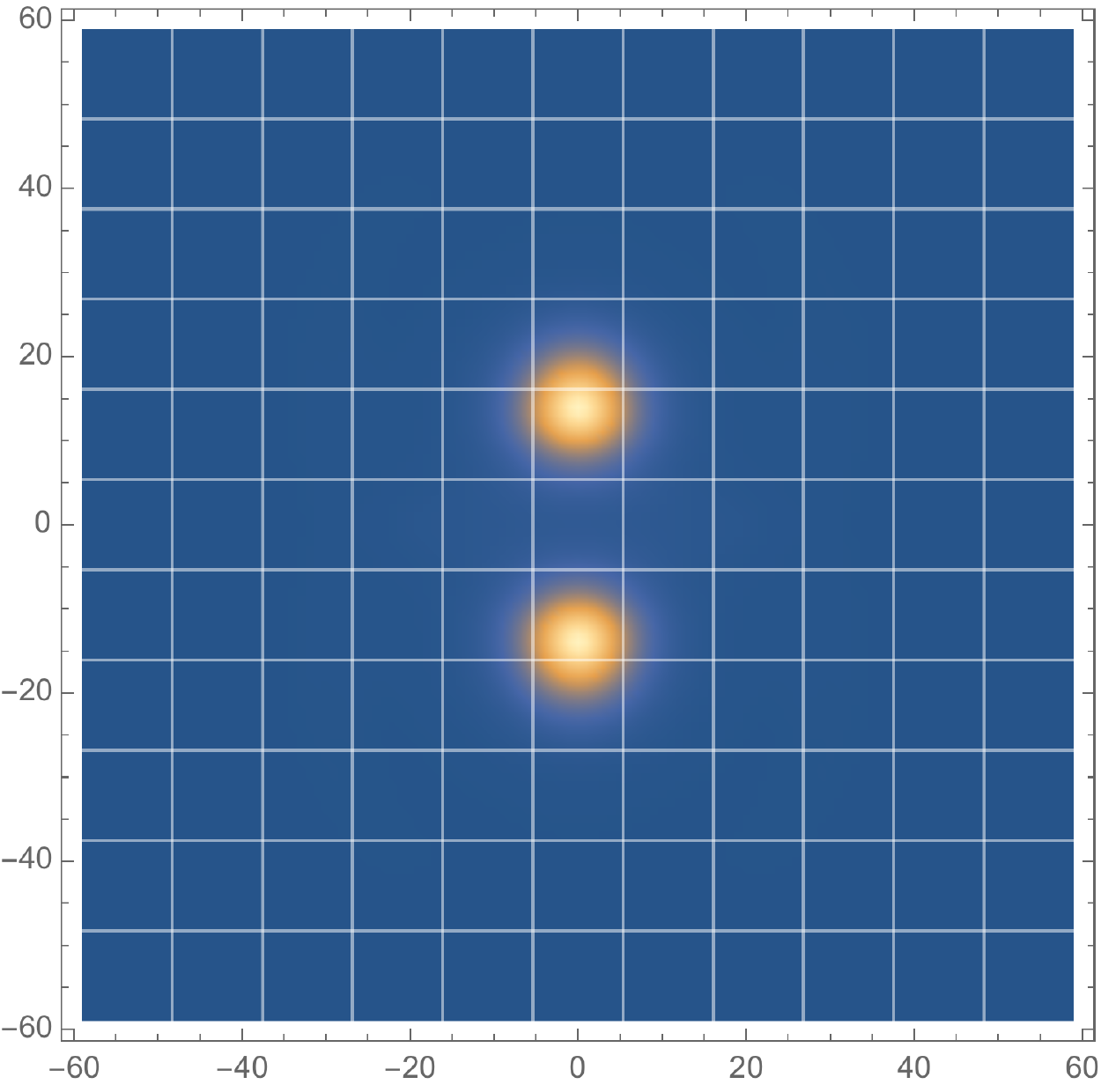}
  \includegraphics[height=0.3\textwidth,angle=0]{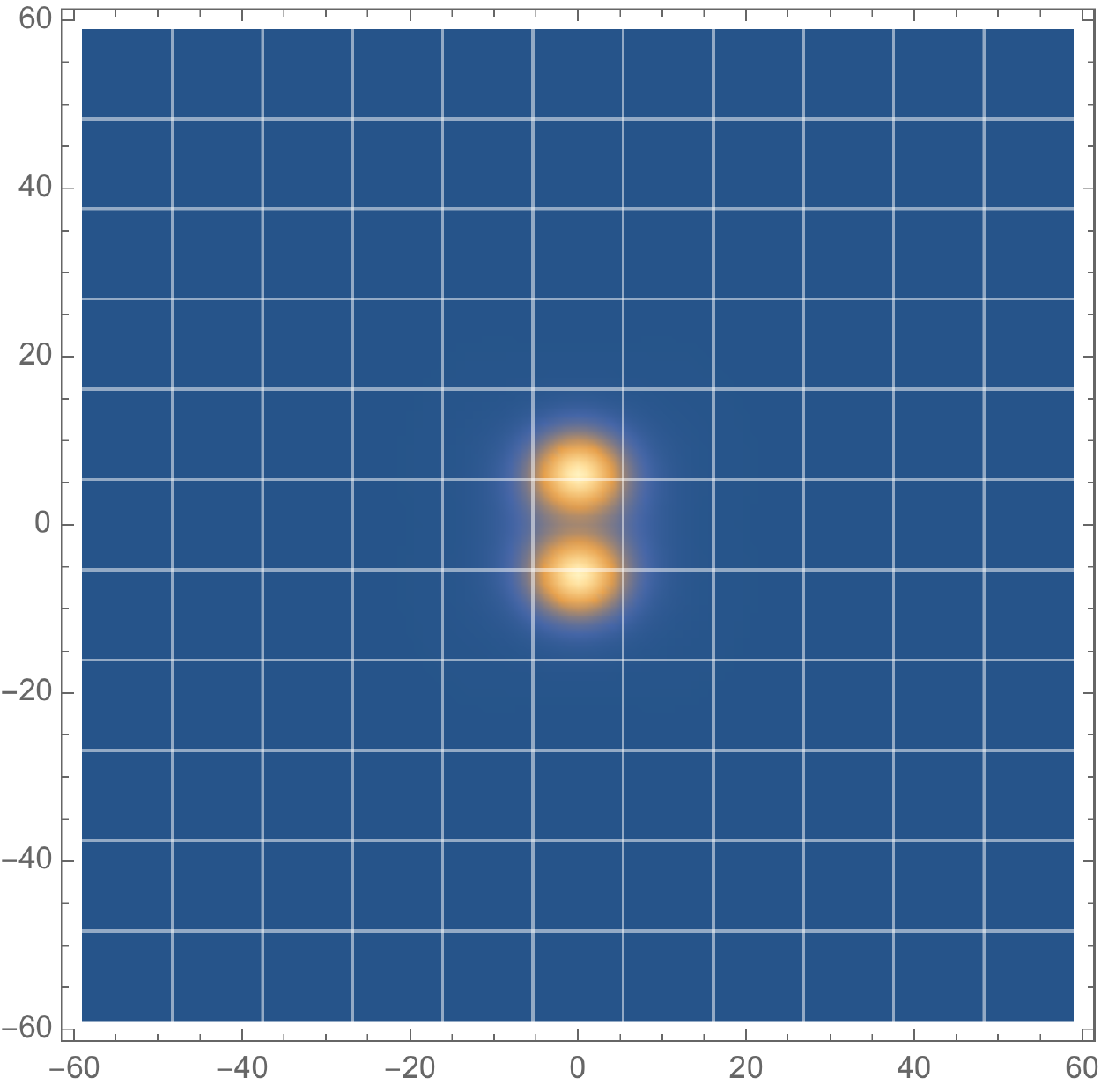}
  \caption{Snapshots of a planar slice of non-annihilating monopole and antimonopole for $\lambda=1$,
  $\gamma=\pi$, and $v_z=0.5$. Except for the twist, all parameters, including snapshot times, are 
  identical to those in Fig.~\ref{snapshots1}. The colors represent energy density. At yet later
  times, the monopoles back-scatter but are still bound and return to annihilate as discussed in the text.}
\label{snapshots2}
\end{figure}

\begin{figure}
\begin{center}
  \includegraphics[height=0.35\textwidth,angle=0]{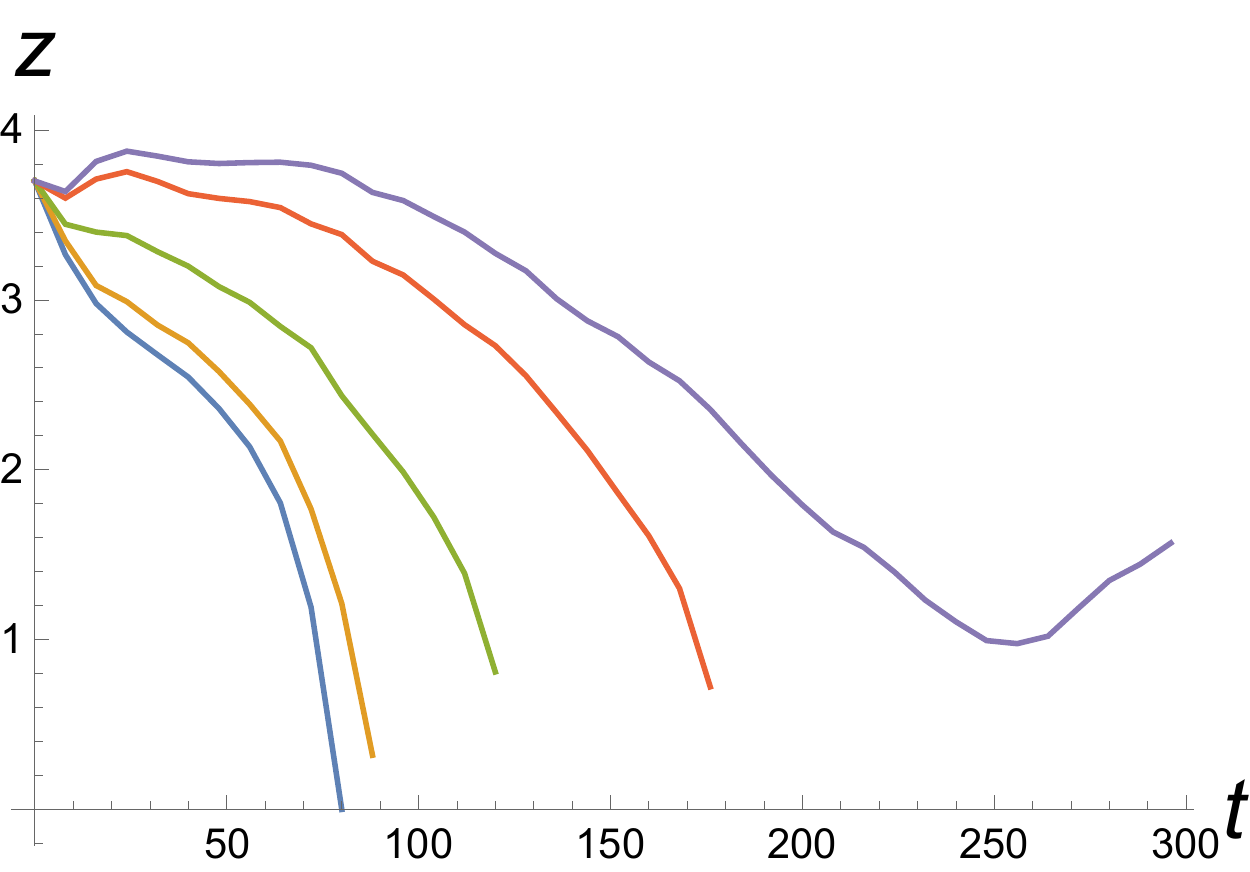}
\end{center}
  \caption{The z-coordinate of the monopole as a function of time for $\lambda=1$,
  $v_z = 0.50$  and $\gamma/(\pi/4) =0, 1, 2, 3, 4$ (curves from left to right).
  The curves terminate once ${\rm min}[\sqrt{\phi^a\phi^a}] \ge 0.25$ (a condition that
  is met after the \mmbar have annihilated) except in the $\gamma=\pi$
  case, when the \mmbar have not annihilated even by the end of the simulation 
  run (300 time steps with $dt=dx/2=0.1$).}
\label{zvstforgamma}
\end{figure}

The initial conditions for a monopole-antimonopole pair were also evolved
using the classical equations of motion in Ref.~~\cite{tanmayscattering}. 
Fig.~\ref{snapshots1}
shows snapshots of the scattering for an untwisted \mmbar, while Fig.~\ref{snapshots2}
shows the scattering when the \mmbar are twisted ($\gamma =\pi$). In the untwisted
case, the \mmbar annihilate, while in the twisted case they come close but not
close enough to annihilate. In fact, they bounce back as seen in Fig.~\ref{zvstforgamma}.

\subsection{$M{\bar M}$ Creation}
\label{mmbarcreation}

Next we discuss the creation of $M{\bar M}$. 
The issue is that monopoles are described as solutions to the classical equations
of motion, while particles are quanta in a quantum field theory. So \mmbar
creation requires a description of magnetic monopoles in a quantum field theory.
Further, a monopole is a non-perturbative solution since its magnetic charge
is proportional to $1/g$ where $g$ is the gauge coupling. 
Perturbative calculations can only give a power law expansion in $g$ and
cannot describe monopoles. Thus any attempt at a perturbative calculation of the 
\mmbar creation amplitude is sure to fail.

We take a different approach to \mmbar creation. We do not consider \mmbar
creation from the collision of just two (or few) particles. Instead we consider the 
collision of classical wavepackets of gauge particles. These classical wavepackets
contain a large number of particles. With such initial conditions, it becomes possible
to study \mmbar creation because we can simply evolve the initial conditions using
the classical equations of motion. However, even within this classical framework, it is not
clear what initial conditions will lead to \mmbar creation and it requires some
physical intuition, guesswork and luck to find initial conditions that successfully
produce monopoles.

If two wavepackets of gauge fields with sufficient energy collide, we can expect that 
a monopole-antimonopole pair can be created but this is not sufficient 
because we also require that the monopole and antimonopole separate from 
each other and do not re-annihilate. On the other hand, a monopole and 
an antimonopole
attract each other by the Coulomb force and this will tend to bring them
together, causing them to annihilate. However, if the \mmbar
are twisted, as discussed in the previous sections, the twist could 
provide a repulsive force between the monopole and antimonopole and
may help separate them. 
This suggests that the initial gauge wavepackets carry some parity
violating structure, which is possible if they are circularly polarized.

Some more intuition may be gained from a result in
magneto-hydrodynamics (MHD), that the magnetic helicity is a
conserved quantity. 
For our purposes, the magnetic helicity can
be thought of as a measure arising due to twisted or helical magnetic 
field lines. 
Further, magnetic helicity likes to spread out. 
Therefore, if one provides initial conditions that compress magnetic helicity, 
the system will try and resist. But the only way out is to break the MHD
approximation and one way for this to happen is by the creation of
magnetic monopoles. So perhaps the compression of magnetic helicity
can lead to \mmbar creation.

After this guesswork, the following circularly polarized gauge wavepackets
were used to construct the initial conditions:
\begin{eqnarray}
{\cal W}^3_x &=& \partial_y f_1 [ (\omega  f_2  - \partial_z f_2 ) \cos(\omega (t+(z-z_0))) -
 (\omega '  f_3 + \partial_z f_3 )\cos (\omega' (t-(z+z_0))) ] 
\nonumber  \\
{\cal W}^3_y &=& \partial_x f_1 [ (\omega f_2  +\partial_z f_2 ) \sin(\omega (t+(z-z_0))) -
(\omega ' f_3 -\partial_z f_3 )\sin(\omega '(t-(z+z_0))) ]
\nonumber  \\
{\cal W}^3_z &=& \partial_x\partial_y f_1 [ f_2 \{ \cos(\omega (t+(z-z_0))) -\sin(\omega (t+(z-z_0))) \} 
 \nonumber \\ && \hskip 3 cm
 +  f_3  \{ \cos(\omega' (t-(z+z_0)))]-\sin(\omega ' (t-(z+z_0))) \} ]
\end{eqnarray}
where the profile functions $f_1$, $f_2$ and $f_3$ will be specified shortly.
Note that this is not a solution to the field equations. It is simply a
configuration that represents a gauge wavepackets that is moving in
the $-z$ direction. The configuration is complicated due to the requirement that the
initial conditions need to satisfy the Gauss constraints. The chosen configuration
satisfies $\nabla \cdot {\bm W}^3=0$, and the electric field ${\bm E}^3 = -\partial_t {\bm W}^3$
satisfies the Gauss constraint with vanishing charge density. We will arrange
for an initially vanishing charge density by setting $\partial_t \phi^a |_{t=0} =0$ when we
choose initial conditions for the scalar field.

The profile functions are chosen to create localized packets in all directions
\begin{equation}
f_1(x,y) = a~\exp \left [ - \frac{(x^2+y^2)}{2w^2} \right ]
\end{equation}
\begin{equation}
f_{2,3}(t\pm (z\mp z_0)) = \exp \left [ - \frac{(t\pm (z\mp z_0))^2}{2w^2} \right ]
\end{equation}
where $a$ is an amplitude and $w$ is a width.
The frequencies $\omega$ and $\omega '$ can be different in general but here
we only consider $\omega' = \pm \omega$. The case $\omega '=\omega$
corresponds to scattering of left- and right-handed circular polarizations, while
$\omega'=-\omega < 0$ corresponds to scattering of left- on left-handed 
circular polarization waves. 

Now we can state our initial conditions for the gauge field,
\be
W^a_i (t=0,{\bf x}) = {\cal W}^a_i (t=0,{\bf x}), \ \ 
\partial_t W^a_i (t=0,{\bf x}) = \partial_t {\cal W}^a_i (t=0,{\bf x})
\ee
For the scalar field we choose
\be
\phi^a (t=0,{\bf x}) = \frac{\eta}{\sqrt{2}} (1,0,1), \ \
\partial_t\phi^a (t=0,{\bf x}) = 0
\ee

Next we need to evolve these initial conditions.
The SO(3) equations of motion are written in a form inspired by Numerical Relativity
that provides for numerical stability~\cite{baumgarte2010book}.
The field theory parameters in the numerical work are $g=0.5$, $\lambda=1$, $\eta=1$,
and the initial condition parameters were chosen to be $w=0.4$, $z_0=1$, $a=10$, 
$\omega=4$, $\omega'=-4$. 

\begin{figure}
\begin{center}
  \includegraphics[height=0.35\textwidth,angle=0]{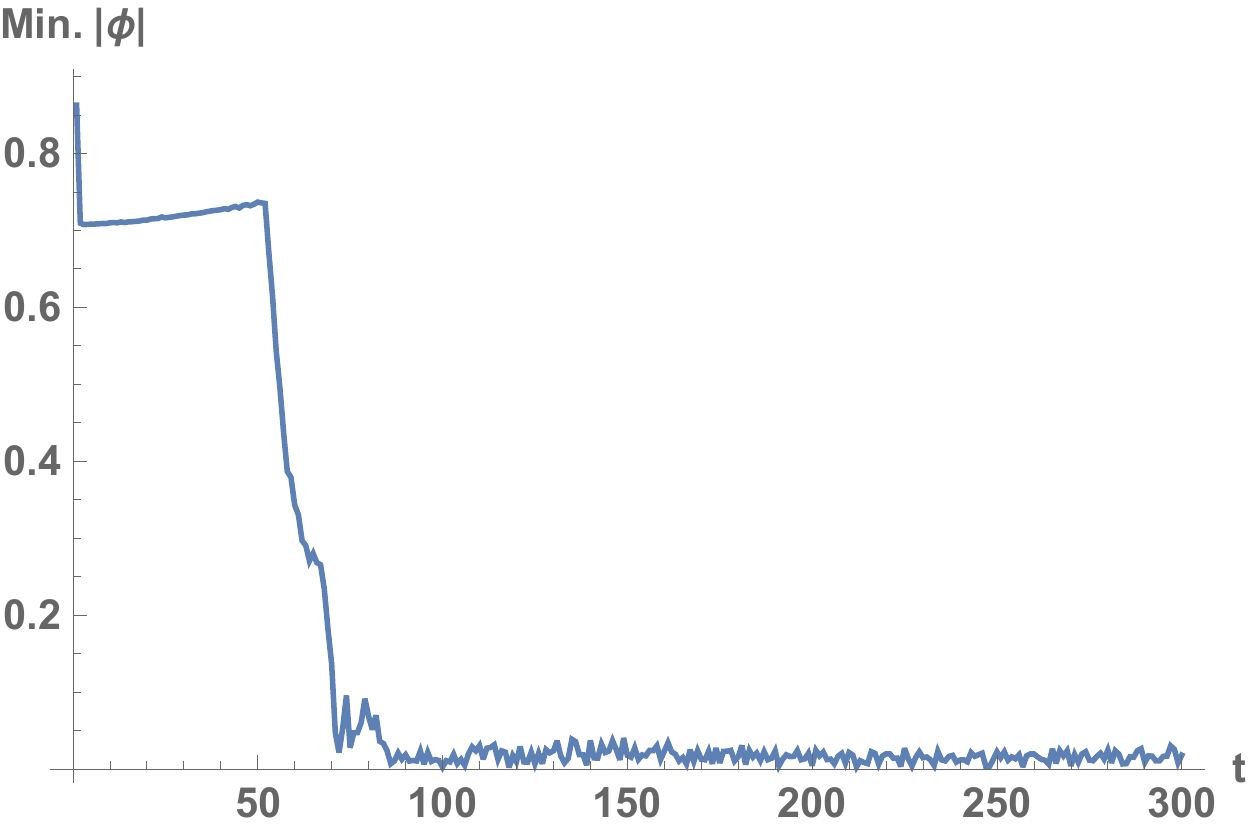}
\end{center}
  \caption{Minimum value of $|{\vec \phi}|$ on the lattice as a function of time showing
  that zeros of the scalar field are produced after some evolution.}
\label{phimin}
\end{figure}

A first sign that monopoles are produced is that the evolution produces zeros of the
Higgs field. In Fig.~\ref{phimin}, the minimum value of $|{\vec \phi }|$ over the entire
simulation box is plotted as a function of time. The sharp drop after some time and
the persistence of the zero value indicates that monopoles have been produced and
survive until the end of the evolution.

\begin{figure}
  \includegraphics[height=0.25\textwidth,angle=0]{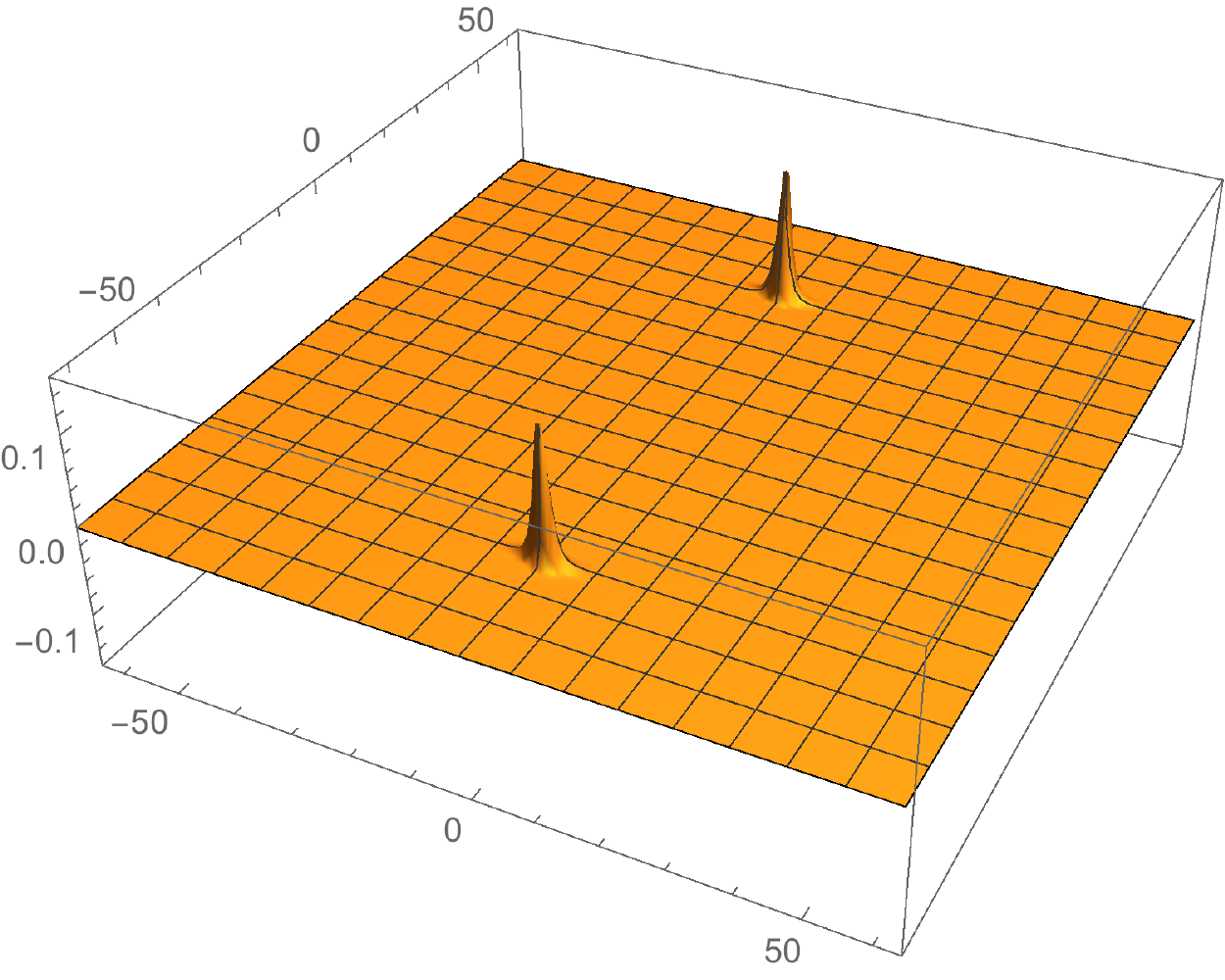}
   \includegraphics[height=0.25\textwidth,angle=0]{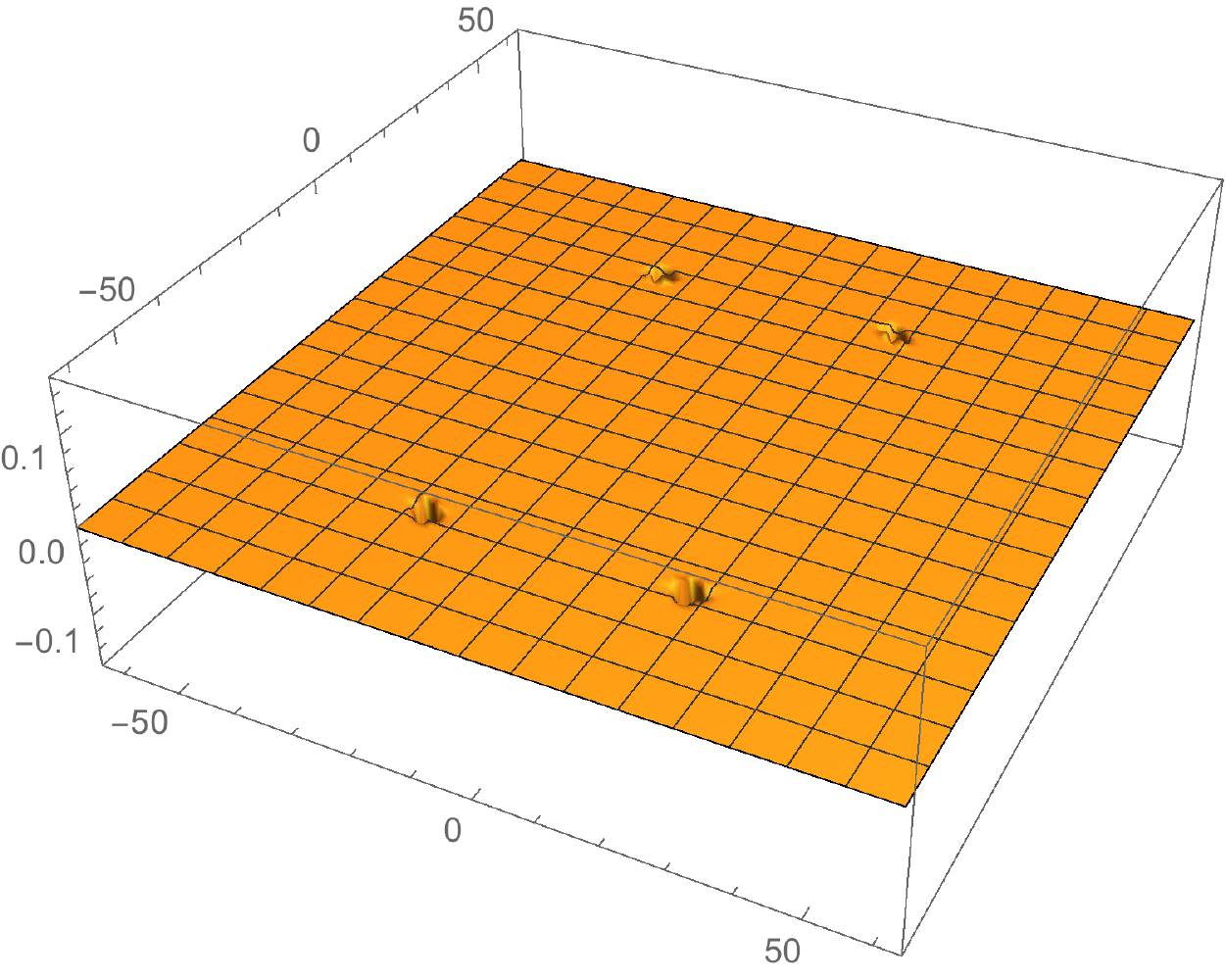}
    \includegraphics[height=0.25\textwidth,angle=0]{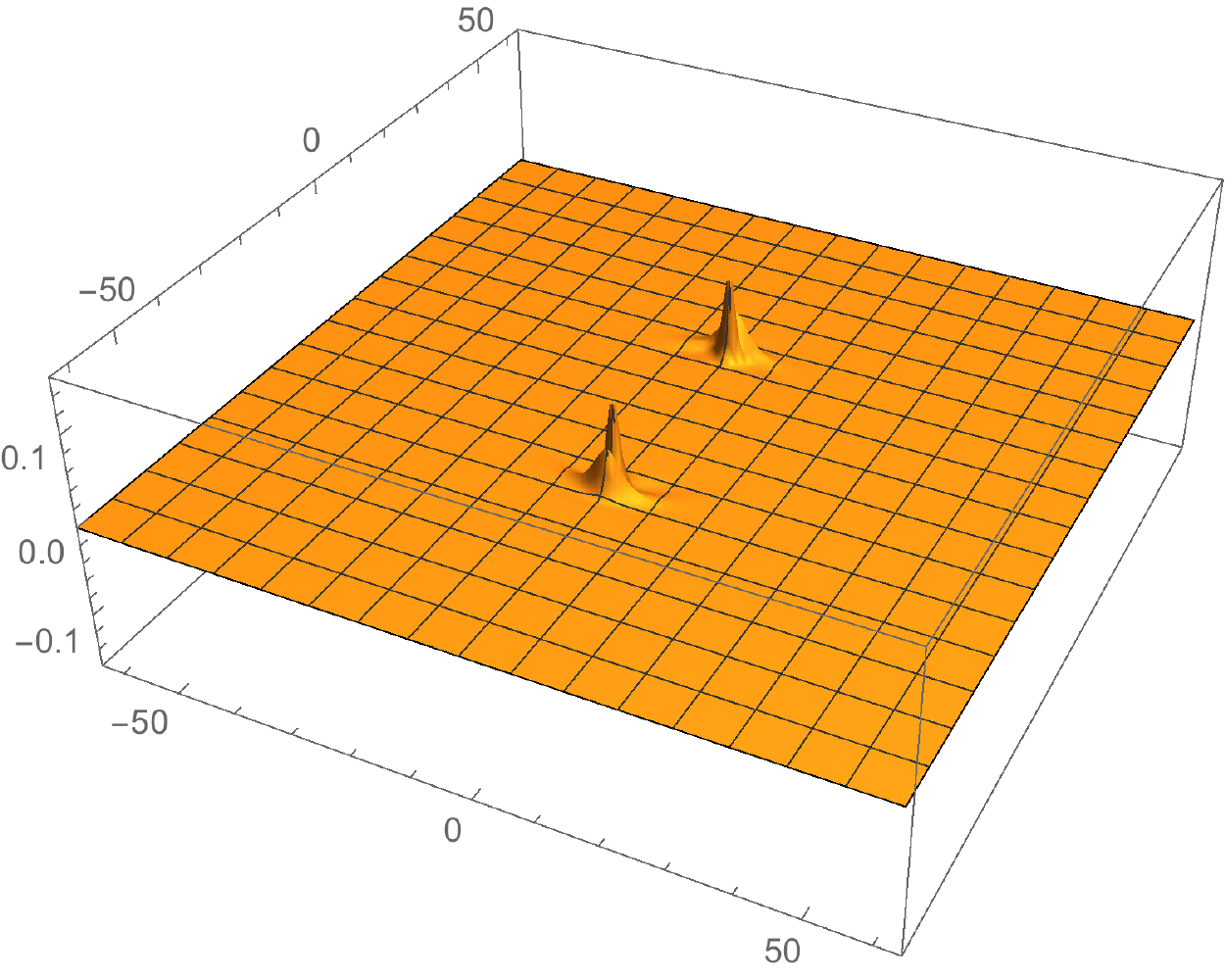}
  \caption{Topological winding at late times on slices with 
  $z=2.9,~3.7$ and $5.7$ for simulations on a $128^3$ lattice
  with  $dx=0.1$ and $z=0$ at the center of the lattice.
  The total topological charges on these slices are +2, -4, and +2 respectively.}
\label{windingonslices}
\end{figure}

Another characteristic of magnetic monopoles is that they are topological structures
and so there is a topological charge density that should be non-vanishing at the
location of the monopoles. The formula for the topological charge within a surface $S$
is given by,
\begin{eqnarray}
W(S) &=& 
\frac{1}{8\pi} \oint_S d{\hat n}^i 
\epsilon_{ijk}\epsilon_{abc} {\hat \phi}^a \partial_j{\hat \phi}^b\partial_k{\hat \phi}^c 
\end{eqnarray}
where ${\hat \phi}^a=\phi^a/|{\vec \phi}|$. Fig.~\ref{windingonslices} shows three
different slices of the simulation box at  $z=2.9,~3.7$ and $5.7$ at the end of
the run. The plots show 4 positive peaks at $z=2.9$ and $5.7$ and 4 negative
peaks at $z=3.7$. Thus the run produces 4 monopoles and 4 antimonopoles.

These numerical experiments are proof of concept showing that a class of initial 
conditions can be constructed to create magnetic monopoles.

\section{String Creation}
\label{stringcreation}

Just as we considered the creation of \mmbar, we can consider the creation of strings (vortices). 
The problem here is that only loops of strings can be created and these will be ephemeral as they
will re-collapse and annihilate. On the other hand, vortices are present in superconductors and
so some of these ideas can be implemented in the laboratory.

String solutions exist in the Abelian Higgs model that is given by the Lagrangian,
\begin{equation}
        \mathcal{L}=-\frac{1}{4} F_{\mu \nu} F^{\mu \nu} + 
        \frac{1}{2}| D_\mu\phi |^2-\frac{\lambda}{4} \left( |\phi |^2-\eta^{2} 
        \right)^{2} 
\end{equation}
where $\phi = \phi_1 + i \phi_2$ is a complex scalar field,
$D_\mu = \partial_\mu + i e A_\mu$,
$A_\mu$ is the U(1) gauge field with field strength tensor
$F_{\mu \nu} = \partial_\mu A_\nu - \partial_\nu A_\mu$, and
$\lambda$ and $e$ are coupling constants.

The solution for a straight string along the $z-$axis is,
\be
\phi = \eta f(r) e^{i\theta}, \ \
A_i = v(r) \epsilon_{ij} \frac{x^j}{r^2} \ \ (i,j=1,2)
\ee
where we work in cylindrical coordinates $r=\sqrt{x^2+y^2}$,
$\theta=\tan^{-1}(y/x)$, $f(r)$ and $v(r)$ are profile functions that vanish
at the origin and go to 1 asymptotically.
The energy per unit length (also the tension) of the string is given by
\be
\mu = \pi \eta^2 F(\beta )
\ee
where $\beta \equiv 2\lambda/ e^2$. The function $F(\beta)$ is known
numerically and is a smooth, slowly varying function with $F(1)=1$,
$m_S = \sqrt{2\lambda} \eta$ equals the vector mass,  $m_V =  e \eta$.
For $\beta$ not too large, the thickness of the scalar fields in the string is
$\sim m_S^{-1}$ and of the vector fields is $\sim m_V^{-1}$. We will
only consider $\beta=1$.

The string is characterized by a topological winding number that is defined by
\be
n = \frac{-i}{2 \pi \eta^2} \oint dx^i \phi^* \partial_i \phi =
\frac{1}{2 \pi} \oint \frac{d \theta}{dl} dl
\label{windingdefn}
\ee
where $\theta$ is the phase of the scalar field at a given point on the
contour and $l$ denotes the parameter along the integration curve.

We based the initial conditions for our simulations on those used for 
monopole-antimonopole production in Sec.~{mmbarcreation} with 
$W^3_\mu$ in the monopole case corresponding to $A_\mu$ in the
string case.
Now the initial conditions for the gauge fields and their time derivatives are,
\ba
A_i (t=0,{\bf x}) &=& {\cal A}_i (t=0,{\bf x}), \\
\partial_t A_i (t=0,{\bf x}) &=& \left [ \partial_t {\cal A}_i (t,{\bf x}) \right ]_{t=0}
\ea

The initial conditions for the scalar field are ``trivial'',
\be
\phi (t=0,{\bf x}) = \eta, \ \ [\partial_t \phi (t,{\bf x}) ]_{t=0} = 0.
\ee

\begin{figure}
\includegraphics[height=0.24\textwidth,angle=0]{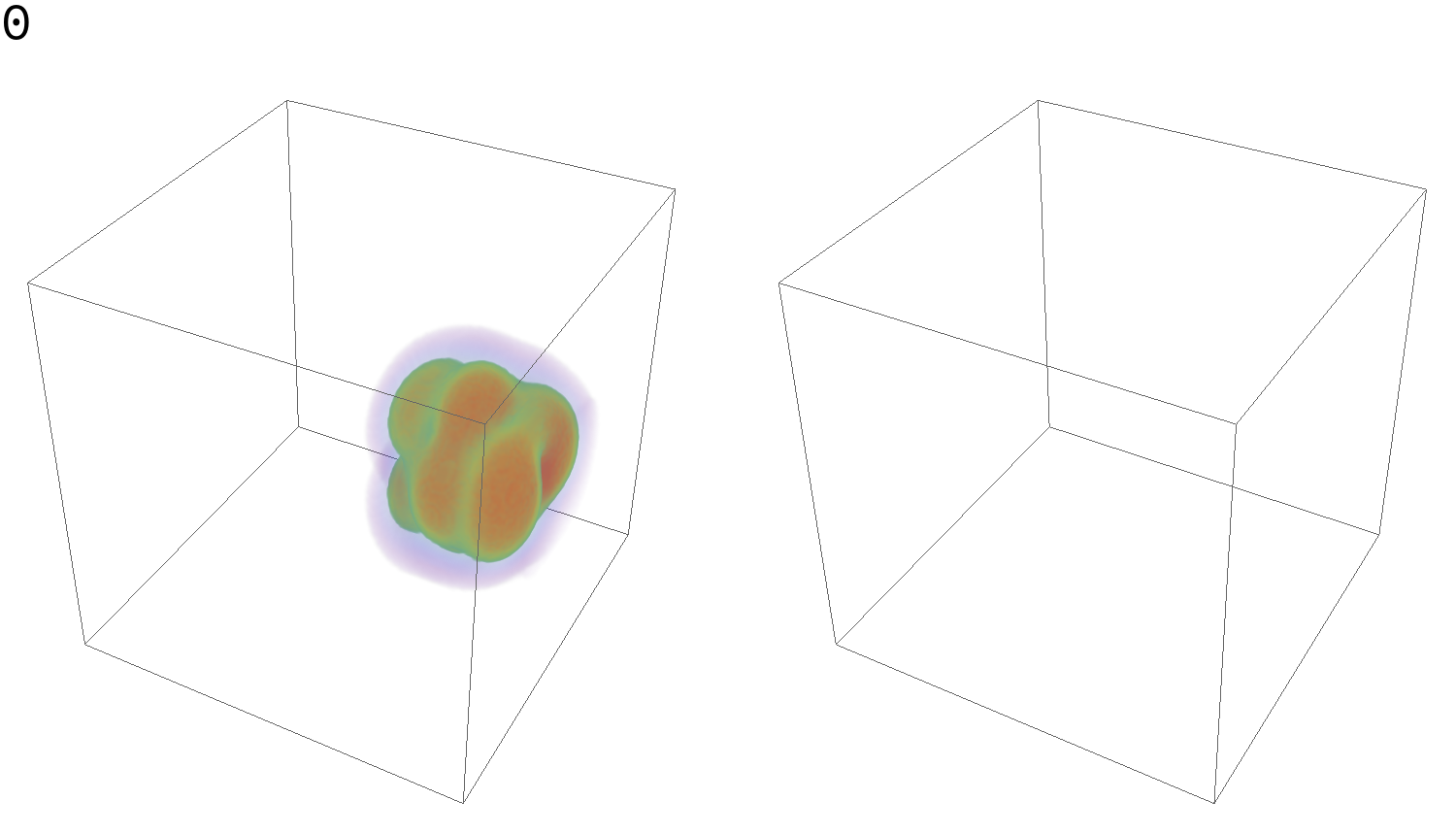} 
\hskip 1.5 cm
\includegraphics[height=0.24\textwidth,angle=0]{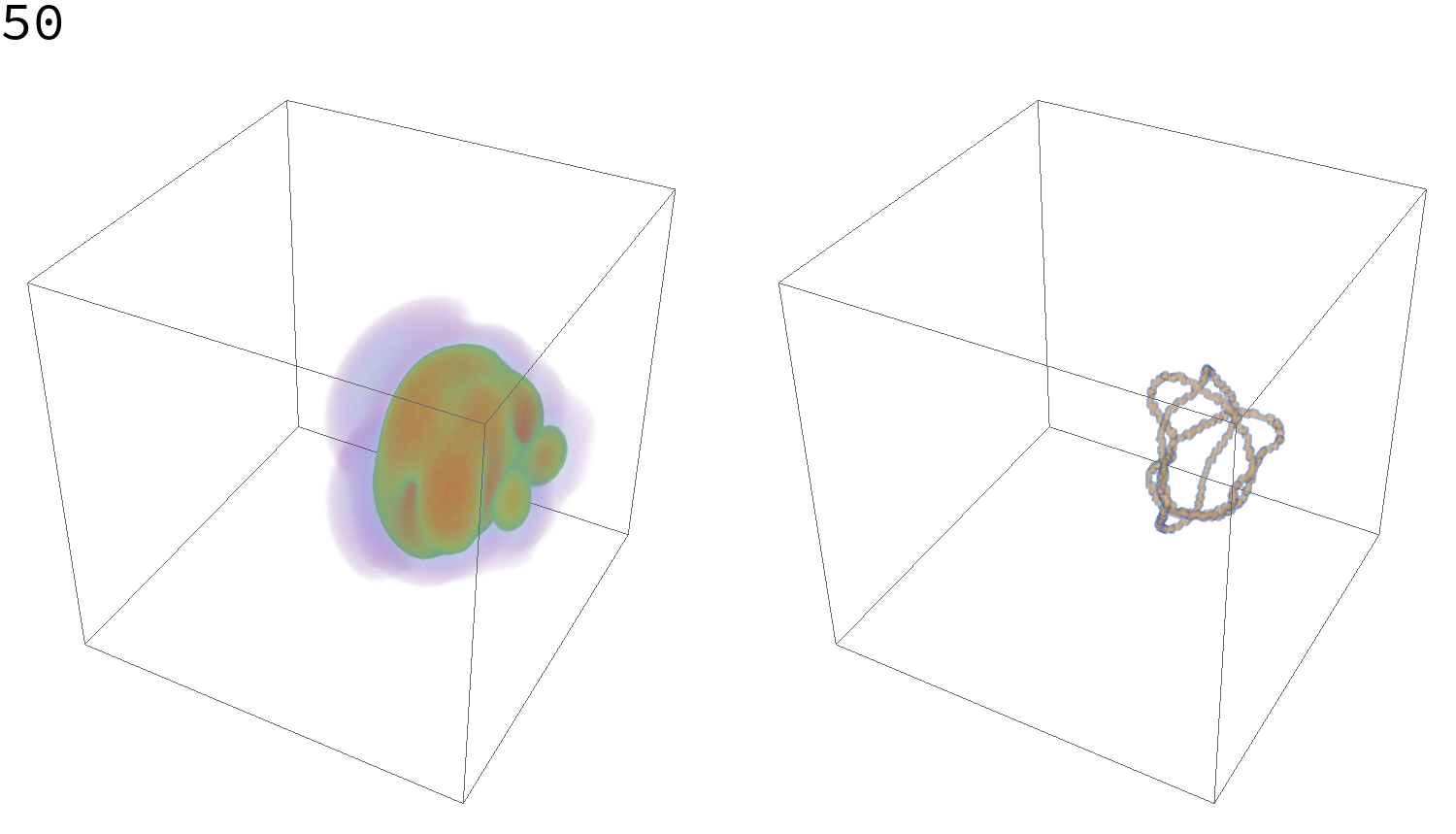}\\
\includegraphics[height=0.24\textwidth,angle=0]{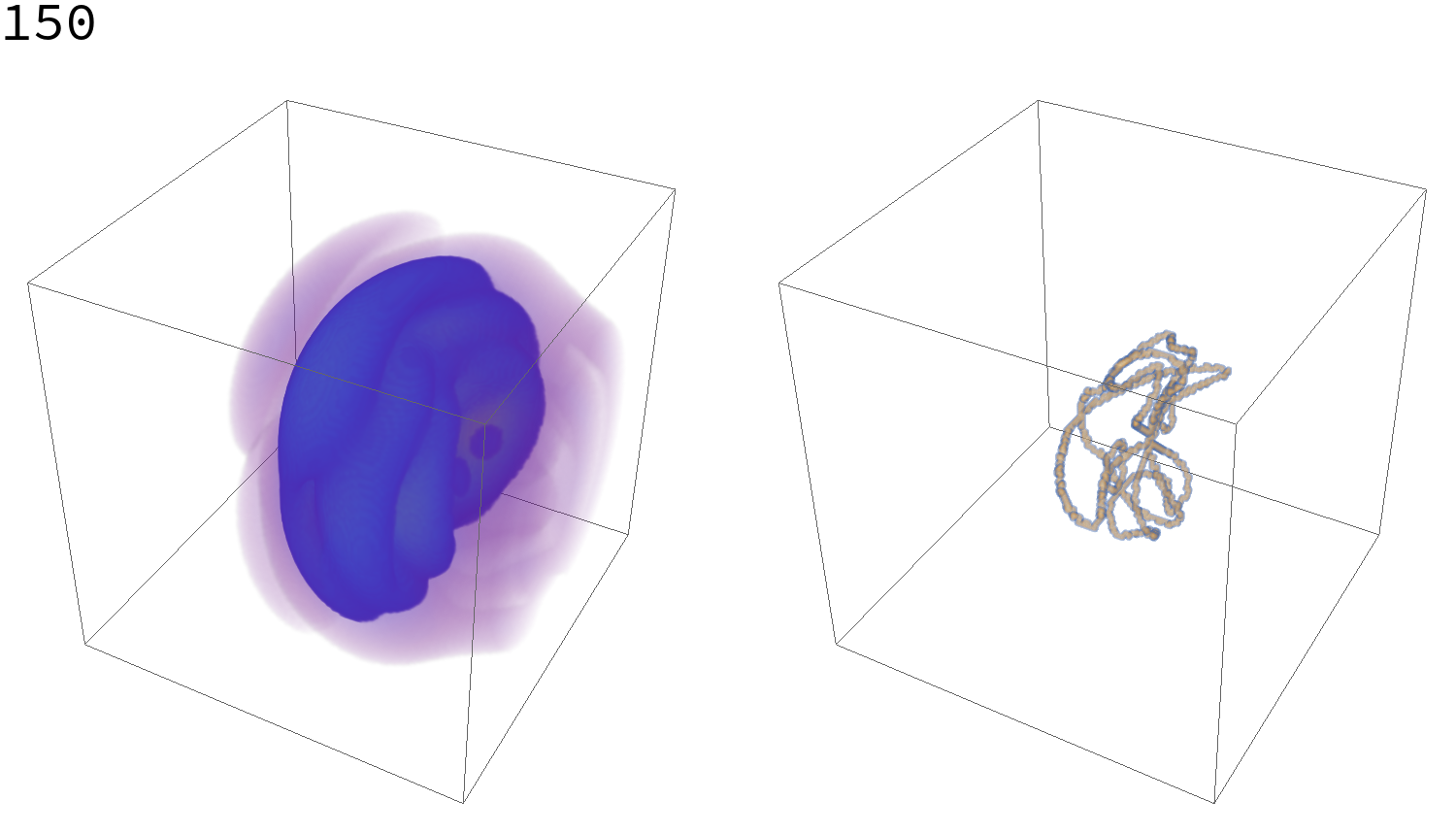} 
\hskip 1.5 cm
\includegraphics[height=0.24\textwidth,angle=0]{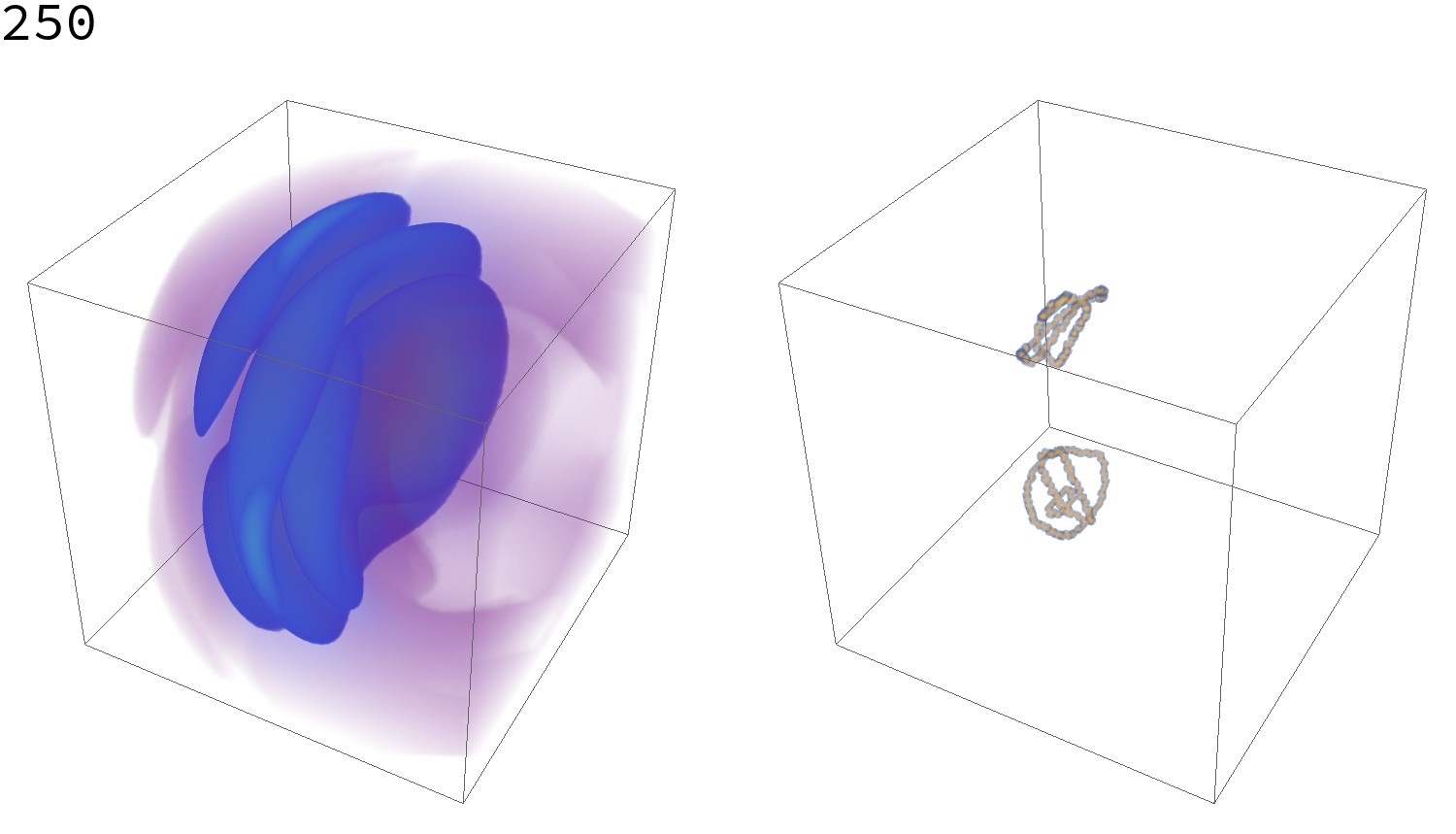} 
\begin{center}
\includegraphics[height=0.07\textwidth,angle=0]{tvAyushFig5}
\end{center}
\caption{Total energy density (boxes on the left) and winding (boxes on the 
		right) at different time steps for the case of one wavepacket
		(``prompt production of strings'') for sample parameters.
		}
\label{SinglePulseMovie}
\end{figure}

\begin{figure}
	\centering
\includegraphics[height=0.40\textwidth,angle=0]{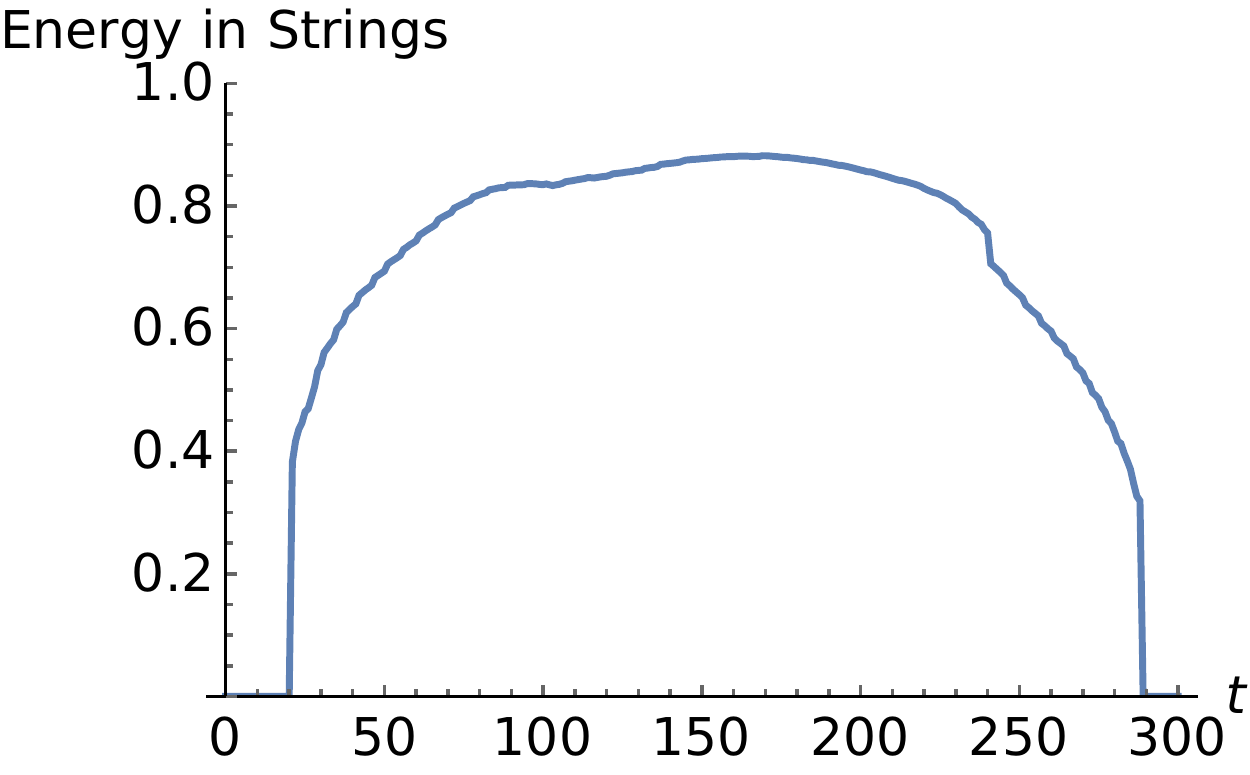}
		\caption{Energy in strings as a fraction of total energy 
		versus time(-steps).
}
 \label{EnergyInStringsSingle}
\end{figure}

With these initial conditions, we were able to 
explore the parameter space for the formation of U(1) gauge strings in two settings: 
(i) the prompt formation of strings from gauge fields
 and 
(ii) the formation of strings when gauge wavepackets collide.  Here we only show the
results of prompt string formation in which a single wavepacket of gauge fields evolves
to produce strings as in Fig.~\ref{SinglePulseMovie}. 

Unlike the case of magnetic monopoles, the string loops that are formed are 
short-lived as they collapse and produce radiation as is evident from 
Fig.~\ref{EnergyInStringsSingle}. The loops may live longer
if we could find initial conditions that provide them with greater angular momentum
but these too will radiate and dissipate. On the other hand, once a magnetic monopole
and antimonopole pair are produced with sufficient velocity, they will move apart and
survive indefinitely. Furthermore, magnetic monopoles are localized objects and
so the colliding wavpackets need not be very extended. For strings, on the other
hand, the wavepackets
have to extend over a region that is the size of the string loop that is to be produced,
and only relatively small loops can be produced.
In these respects it appears that magnetic monopoles are easier to produce than
strings. 

The flip side is that we know systems that contain gauge strings while
the existence of magnetic monopoles is still speculative.
Gauge strings are known to exist in superconductors and, in that setting, our gauge field 
wavepackets correspond to photon wavepackets. This suggests that by shining light on 
superconductors we could produce strings within the superconductor. However, a realistic 
superconductor is described by a different set of equations that take into 
account the dependence of the model parameters on the temperature.  It will be 
interesting to adapt our analysis to study string production in 
superconductors.

\section{\mmbar in Electroweak Model}
\label{electroweak}

The electroweak model is based on the symmetry breaking
\be
[SU(2)\times U(1)]/Z_2 \rightarrow U(1)_{\rm EM}
\ee
In this case, the initial symmetry group is not simply connected and it requires
some care to see that there are no topological monopoles in the model. A simpler 
way to see the absence of topological 
magnetic monopoles is that the electroweak Higgs, $\Phi$, is an $SU(2)$ doublet
\be
\Phi = \begin{pmatrix} \phi_1+i\phi_2 \\ \phi_3+i\phi_4 \end{pmatrix}
\ee
and the minimum of the Higgs potential is given by
\be
\Phi^\dag \Phi = \phi_1^2 + \phi_2^2 + \phi_3^2 + \phi_4^2 = \eta^2
\ee
where $\eta$ is the vacuum expectation value of the Higgs. Thus the vacuum
manifold is a three-sphere that does not admit incontractible two-spheres that
are necessary to obtain topological monopoles.

The absence of topology in the standard model still leaves room for {\it confined}
magnetic monopoles. Indeed it was shown in Ref.~\cite{nambu,Achucarro:1999it} 
that such magnetic monopoles do exist in the electroweak model.
These electroweak monopoles carry magnetic charge but are attached 
to a string made of $Z$ fields that connect the monopole to an antimonopole. 
The situation is very similar to that of a quark that carries electric charge but
is confined by a QCD flux tube that connects it to an anti-quark of opposite
electric charge.

Can we create electroweak monopole-antimonopole pairs by colliding
gauge wavepackets? Even if we manage to create a monopole-antimonopole
pair, they will be confined by the $Z-$string that will pull them together and
cause them to annihilate. So electroweak monopoles can at best be
created temporarily, somewhat like the string loops we discussed in
Sec.~\ref{stringcreation}.

There is one situation which is of interest even if electroweak monopoles
are produced and that then annihilate. This is if the monopoles are produced
but then annihilate after they have twisted by $2\pi$. A signature of such
an event will be a change in the Chern-Simons number which is defined as
\be
CS  = \frac{1}{32\pi^2}\epsilon^{ijk}\int d^3x \biggl [
g^2 \left ( W^a_{ij} W^a_k -\frac{g}{3} \epsilon_{abc} W^a_i W^b_j W^c_k \right ) -g'^2Y_{ij}Y_k \biggr ],
\label{cs}
\ee
where $W^a_\mu$ are the $SU(2)$ gauge fields and $Y_\mu$ the
hypercharge gauge field. Changes in the Chern-Simons number are
also indicative of baryon number violation in the electroweak model
when fermions are included.

The initial conditions of Sec.~\ref{mmbarcreation} used to study monopole creation
in the SO(3) model can be adapted to the electroweak model. Numerical
experiments as of this time have not yielded a successful Chern-Simons
number changing event but the search continues.

\section{Conclusions}
\label{conclusions}

Magnetic monopoles are predicted in a wide class of particle physics models,
{\it e.g.} all models of Grand Unification. Indeed, the realization that Grand
Unification and standard cosmology predicts an over-abundance of magnetic 
monopoles led to the proposal of inflationary cosmology that vastly dilutes
the monopole abundance. Thus monopoles are expected to be present in
the physics that governs the universe but are not realized in our observable
universe. 

This peculiar circumstance is not special to magnetic monopoles. The underlying
physics also admits other structures such as cars and computers but these do
not occur naturally. Instead they require human intervention for their existence.
Magnetic monopoles fall in this category -- they may require humans to create
them. Whether humans have the will to create monopoles is another question
and the answer will hinge on their perceived utility. (One potential utility is to
use magnetic monopoles for catalyzing proton decay and to harness the
released energy.)

These considerations have led us to study the interactions and dynamics of
magnetic monopole-antimonopole pairs. The interaction of \mmbar is non-trivial
because of a twist that was known to exist from the 70's~\cite{Taubes:1982ie,Taubes:1982ie2}. 
We have described numerical work that confirms the picture and quantifies it more
accurately. The twist also leads to non-trivial dynamics. The scattering of
monopole-antimonopole can lead to a bounce instead of annihilation. We
have used these results to intuit initial conditions that can lead to monopole
creation and have successfully tested them to see the production of four
monopoles and four antimonopoles. Similar studies that lead to the production
of strings may be relevant to superconductors in the lab. When these
methods are applied to the electroweak model, they can potentially
generate Chern-Simons number changing events that lead to baryon
number violation. However we have not yet seen such an event in
our numerical experiments and are continuing our explorations.

\enlargethispage{20pt}

%\ethics{Insert ethics statement here if applicable.}

%\dataccess{Insert details of how to access any supporting data here.}

\aucontribute{
AS and TV developed the numerical code, performed the analysis and drafted the manuscript.
%For manuscripts with two or more authors, insert details of the authors’ contributions here. This should take the form: 'AB caried out the experiments. CD performed the data analysis. EF conceived of and designed the study, and drafted the manuscript 
All authors read and approved the manuscript.}

\competing{The author(s) declare that they have no competing interests.}

\funding{Insert funding text here.}

\ack{AS and TV are supported by the U.S. Department of Energy, Office of High Energy 
Physics, under Award No.~DE-SC0018330 at Arizona State University.}

%\disclaimer{Insert disclaimer text here if applicable.}

%%%%%%%%%% Insert bibliography here %%%%%%%%%%%%%%

%\bibstyle{aps}
\newpage
\bibliographystyle{RS}
\bibliography{paper}

\end{document}